\documentclass[12pt,showkeys,amsmath,amssymb]{revtex4}
\usepackage{amsmath,amsfonts,amsthm,amscd,amssymb,latexsym}
\usepackage{bm}
\usepackage{dcolumn}
\usepackage{graphicx}
\usepackage{epstopdf}
\usepackage{color}
\usepackage{epsf}
\usepackage{epsfig}
\usepackage{graphicx, epic, eepic, color}
\usepackage{soul}
\usepackage{tikz}
\usetikzlibrary{calc}
\usetikzlibrary{decorations.markings}
\usetikzlibrary{shadings, calc, decorations.markings}

\newcommand{\beq}{\begin{equation}}
\newcommand{\eeq}{\end{equation}}

\def\@{\partial_}

\def\negenspace{\kern-1.1em}

\def\sqr#1#2{{\vcenter{\hrule height.#2pt\hbox{\vrule width.#2pt
height#1pt \kern#1pt \vrule width.#2pt}\hrule height.#2pt}}}

\begin{document}

\title{Dynamical Friction in Nonlocal Gravity}

\author{Mahmood \surname{Roshan}$^{1,2}$}
\email{mroshan@um.ac.ir}
\author{Bahram \surname{Mashhoon}$^{2,3}$}
\email{mashhoonb@missouri.edu}

\affiliation{$^1$Department of Physics, Faculty of Science, Ferdowsi University of Mashhad, P.O. Box 1436, Mashhad, Iran \\
$^2$School of Astronomy, Institute for Research in Fundamental Sciences (IPM), P. O. Box 19395-5531, Tehran, Iran\\
$^3$Department of Physics and Astronomy, University of Missouri, Columbia, Missouri 65211, USA\\
}

\begin{abstract}
We study dynamical friction in the Newtonian regime of nonlocal gravity (NLG), which is a classical nonlocal generalization of Einstein's theory of gravitation. The nonlocal aspect of NLG simulates dark matter. The attributes of the resulting effective dark matter are described and the main physical predictions of nonlocal gravity, which has a characteristic lengthscale of order 1 kpc, for galactic dynamics are presented. Within the framework of NLG, we derive the analogue of Chandrasekhar's formula for dynamical friction. The astrophysical implications of the results for the apparent rotation of a central bar subject to dynamical friction in a barred spiral galaxy are briefly discussed. 
\end{abstract}


\maketitle

\section{Introduction}

Imagine the scattering state of a Newtonian two-body system. Under the Newtonian inverse-square law of gravity, the bodies deviate from their original paths during the scattering process, but  the mechanical system is conservative and the overall positive energy is conserved. Let us choose one of the bodies as the reference and focus attention on its initial momentum vector. In the scattering process, the component of the final deflected momentum of the reference body along the direction of its initial momentum does not change to first order in the gravitational coupling constant $G$, but \emph{decreases} beyond the linear order. The net loss of momentum of the reference particle along its initial direction of motion is the source of the dynamical friction force. Normally, dynamical friction becomes physically significant when an astronomical body of mass $M$ moves relatively slowly through a population of stars of average mass $m \ll M$. It was first calculated in a collisionless gravitational system by Chandrasekhar~\cite{Chandra}. For the applications of dynamical friction in galactic dynamics, see~\cite{Binney}. The main purpose of this paper is to study dynamical friction within the framework of nonlocal gravity theory and derive the analogue of Chandrasekhar's formula in the Newtonian regime of NLG. We now turn to a brief introduction of NLG and its Newtonian regime. 

In the special and general theories of relativity, physics is local~\cite{Einstein}. For instance, the standard approach for extending relativity theory to an accelerated system in Minkowski spacetime involves the pointwise application of Lorentz transformations, which makes physical sense if the velocity of the accelerated system is in effect locally uniform during an elementary act of measurement. The locality assumption is certainly valid for pointlike coincidences of classical point particles and rays of radiation. However, wave phenomena are intrinsically nonlocal by the Huygens principle; moreover, such nonlocality extends to the measurement of the electromagnetic field. To determine the frequency content of an incident wave packet, for example, an accelerated observer must in general employ instantaneous Lorentz transformations over an extended period of time, during which the state of the observer varies continuously.  At the same time, there are invariant local acceleration scales of length and time associated with the accelerated system. Therefore, accelerated systems are in general nonlocal and the past history of accelerated motion must be taken into account~\cite{Mashhoon:1993zz}, thus leading to nonlocal special relativity~\cite{Mashhoon:2008vr}. The deep connection between inertia and gravitation, as revealed in Einstein's development of general relativity~\cite{Einstein}, suggests that gravity should be nonlocal as well. The most natural approach to a nonlocal gravity theory would be to introduce history dependence via a nonlocal constitutive relation in close analogy with the nonlocal electrodynamics of media~\cite{Jackson, L+L, HeOb}. Thus, nonlocal gravitational field equations would reflect an influence (``memory") from the past that endures. History dependence could in fact be a natural feature of the universal gravitational interaction. 

To implement this idea, the first step would involve expressing Einstein's general relativity (GR) in a form that resembles Maxwell's electrodynamics. This can be simply accomplished by the introduction of a preferred frame field in spacetime and employing an extended geometric framework known as teleparallelism. Indeed, there is a well-known teleparallel equivalent of general relativity (TEGR), which is the gauge theory of the group of spacetime translations~\cite{Cho}. Therefore, TEGR, though nonlinear, is formally analogous to electrodynamics and can be rendered nonlocal via history-dependent constitutive relations as in the nonlocal electrodynamics of media~\cite{Hehl:2008eu, Hehl:2009es}. In the resulting theory of nonlocal gravity (NLG), the gravitational field is locally defined but satisfies partial integro-differential field equations. The only known exact solution of NLG is the trivial solution; that is, we simply recover Minkowski spacetime in the absence of gravity. Thus far, the nonlinearity of NLG has prevented finding exact solutions for strong-field regimes such as those involving black holes or cosmological models~\cite{Bini:2016phe}. However, linearized NLG and its Newtonian limit have been the subject of extensive investigations~\cite{Blome:2010xn, Rahvar:2014yta, Mashhoon:2014jna, Chicone:2015coa, MaHe}. A comprehensive account of NLG is contained in~\cite{BMB}.  

When nonlocal TEGR is expressed as modified GR, the source of gravity turns out to include besides the standard symmetric  energy-momentum tensor $T_{\mu \nu}$ of matter, certain purely nonlocal gravity terms which we interpret in terms of nonlocally induced effective dark matter. That is,  the nonlocal aspect of gravity appears to simulate dark matter.  What is now considered dark matter in astrophysics and cosmology may indeed be the manifestation of the nonlocal component of the gravitational interaction. This circumstance constitutes a significant observational consequence of NLG and has been explored in connection with  the gravitational physics of nearby spiral galaxies and clusters of galaxies~\cite{Rahvar:2014yta}. Though much remains to be done, thus far NLG appears to be consistent with observational data regarding the solar system as well as nearby galaxies and clusters of galaxies. However, there are still many issues regarding the astrophysical and cosmological implications of nonlocal gravity that must be resolved~\cite{Chicone:2015sda, Chicone:2017oqt, Roshan:2019xda, Ghafourian:2020uae, Roshan:2021mfc}.

\subsection{Newtonian Regime of NLG}

In this paper, we are interested in the Newtonian regime of nonlocal gravity. In the Newtonian limit, the basic equations of nonlocal gravity simply reduce to the nonrelativistic gravitational force
\begin{equation}\label{I0}
\mathbf{F}(\mathbf{x}) = - m \nabla\Phi(\mathbf{x})\,
\end{equation}
on a test particle of inertial mass $m$ in the gravitational potential $\Phi(\mathbf{x})$, which satisfies the nonlocal Poisson equation~\cite{BMB}
\begin{equation}\label{I1}
\nabla^2\Phi (\mathbf{x}) + \int \chi(\mathbf{x}-\mathbf{y}) \nabla^2\Phi (\mathbf{y})\,d^3y = 4\pi G\,\rho (\mathbf{x})\,,
\end{equation}
where $\chi$ is the constitutive kernel in the Newtonian regime. Here, $\chi$ is assumed to be a universal function independent of the potential; moreover, $\chi$ is a smooth function with certain reasonable mathematical properties which make it possible to write Fredholm integral Eq.~\eqref{I1} in the reciprocal form~\cite{Chicone:2011me}  
\begin{equation}\label{I2}
4\pi G\,\rho (\mathbf{x}) +  \int q(\mathbf{x}-\mathbf{y}) [4\pi G\, \rho(\mathbf{y})]\,d^3y = \nabla^2\Phi (\mathbf{x})\,,
\end{equation}
where $q$ is the reciprocal kernel. Equation~\eqref{I2} can be written as
\begin{equation}\label{I3}
\nabla^2\Phi = 4\pi G\,(\rho+\rho_D)\,, \qquad \rho_D(\mathbf{x})=\int q(\mathbf{x}-\mathbf{y}) \rho(\mathbf{y})\,d^3y\,,
\end{equation}
where $\rho_D$ is the density of effective dark matter and is given by the convolution of the reciprocal kernel $q$ with the density of matter $\rho$.  It follows from the convolution theorem for Fourier integral transforms that in Fourier space $\hat{\rho}_D(\mathbf{k}) = \hat{q}(\mathbf{k})\, \hat{\rho}(\mathbf{k})$, where we define $\hat{s} (\boldsymbol{\xi})$ to be the Fourier integral transform of a suitable function $s(\mathbf{x})$ such that 
\begin{equation}\label{I3a}
\hat{s} (\boldsymbol{\xi}) =  \int s(\mathbf{x})\, e^{-i\,\boldsymbol{\xi} \cdot \mathbf{x}}\, d^3x\,, \qquad 
s(\mathbf{x})=\frac{1}{(2\pi)^3}\,\int \hat{s} (\boldsymbol{\xi})\,e^{i\,\boldsymbol{\xi} \cdot \mathbf{x}}\, d^3\xi\,.
\end{equation}
We note that $\rho_D=0$ if $\rho=0$; hence, there is no effective dark matter in the complete absence of matter. Furthermore, for the sake of simplicity, we have suppressed the possible dependence of $\Phi$, $\rho$ and $\rho_D$ upon time $t$. There is no retardation in the Newtonian regime; hence, $\chi$ and $q$ have no temporal dependence and gravitational memory is purely spatial in the Newtonian limit.

In NLG, we determine the reciprocal kernel $q$ on the basis of observational data. Two simple spherically symmetric functions have been considered in detail, namely, 
\begin{equation}\label{I4}
 q_1 (r) = \frac{1}{4\pi \lambda_0} \,\frac{1+\mu_0\, (a_0+r)}{r\,(a_0 + r)}\,e^{-\mu_0\,r}\,
\end{equation}  
and
\begin{equation}\label{I5}
 q_2 (r) = \frac{1}{4\pi \lambda_0} \,\frac{1+\mu_0\, (a_0+r)}{(a_0 + r)^2}\,e^{-\mu_0\,r}\,,
\end{equation}
where $r = |\mathbf{x} - \mathbf{y}|$. Here, $\lambda_0 \sim 1$~kpc is the basic length scale of NLG, while $a_0$ and $\mu_0$ moderate the short and long distance behaviors of $q$, respectively. It turns out that in the Newtonian regime of NLG, we recover the phenomenological Tohline-Kuhn approach to modified gravity~\cite{Toh,Kuhn,Bek}. Indeed, kernels~\eqref{I4} and~\eqref{I5} are appropriate generalizations of the Kuhn kernel $(4\pi \lambda_0\,r^2)^{-1}$ within the framework of NLG~\cite{BMB}.  

Let us note that for $a_0 = 0$, kernels $q_1$ and $q_2$ both reduce to $q_0$ given by
\begin{equation}\label{I6}
 q_0 (r) = \frac{1}{4\pi \lambda_0} \,\frac{1+\mu_0\,r}{r^2}\,e^{-\mu_0\,r}\,,
\end{equation}
where for any finite radial coordinate $r$,  $q_0 > q_1 > q_2$, since $a_0 > 0$. 

For a point mass $m$ located at the origin of spatial coordinates, $\rho(\mathbf{x}) = m \,\delta(\mathbf{x})$, the nonlocal Poisson Eq.~\eqref{I3} reduces to
\begin{equation}\label{I6a}
\nabla^2\Phi(\mathbf{x}) = 4\pi G\,m\, [\delta(\mathbf{x})+ q(\mathbf{x})]\,,
\end{equation}
where $mq(\mathbf{x})$ is the density of dark matter associated with the point mass $m$. Therefore, the net amount of effective dark matter associated with the point mass is given by $m$ times the integral of the reciprocal kernel $q$ over all space. In this connection, it is useful to define
\begin{equation}\label{I7}
\mathcal{E}_i (r) = 4\pi \int_0^r s^2[q_0(s)-q_i(s)]\,ds\,,\qquad i = 1, 2\,,
\end{equation}
where 
\begin{equation}\label{I8}
4\pi \int_0^\infty s^2\,q_0(s)\,ds = \alpha_0\,,\qquad \alpha_0 := \frac{2}{\lambda_0\,\mu_0}\,.
\end{equation}
It is straightforward to show that 
\begin{equation}\label{I9}
\mathcal{E}_1 (r) = \frac{1}{2} \alpha_0\, \zeta_0\,e^{\zeta_0} [E_1(\zeta_0) - E_1(\zeta_0 + \mu_0r)]\,, \qquad \zeta_0 := \mu_0a_0\,,
\end{equation}
\begin{equation}\label{I10}
\mathcal{E}_2 (r) - 2\, \mathcal{E}_1 (r) = -\frac{1}{2} \alpha_0\, \zeta_0\,\frac{r}{r+a_0}e^{-\mu_0\,r}\,,
\end{equation}
where $E_1$ is the \emph{exponential integral function}~\cite{A+S}. It follows that  $\mathcal{E}_1 (r)$ and $\mathcal{E}_2 (r)$ are positive monotonically increasing functions of $r$ that start from zero at $r = 0$ and for $r \to \infty$ approach $\mathcal{E}_1 (\infty) = \tfrac{1}{2}\alpha_0\, \zeta_0\,e^{\zeta_0} E_1(\zeta_0)$ and $\mathcal{E}_2 (\infty) = 2\, \mathcal{E}_1 (\infty)$, respectively. 

The net amount of effective dark matter associated with a point mass $m$ is thus
\begin{equation}\label{I10a}
m_D = m\alpha_0\,w\,,
\end{equation}
where $w = w_i, i = 1, 2$, depending upon whether the reciprocal kernel is chosen to be $q_1$ or $q_2$, respectively. Here, 
\begin{equation}\label{I11}
w_1= 1- \frac{1}{2}\zeta_0\,e^{\zeta_0} E_1(\zeta_0)\,, \qquad w_2 = 1- \zeta_0\,e^{\zeta_0} E_1(\zeta_0)\,
\end{equation}
and $\mathcal{E}_i(\infty) = (1- w_i)\alpha_0$. If we ignore the reciprocal kernel's short distance parameter $a_0$, then $\zeta_0 = \mu_0 a_0$ vanishes and $w = 1$.  Moreover,  $w_1(\zeta_0)$ and $w_2(\zeta_0)$ decrease somewhat from unity with increasing $\zeta_0$; for instance, for $\zeta_0 \in (0, 0.1]$, $w_1 \in (1, 0.9]$ and $w_2 \in (1, 0.8]$; for the graphs of $w_1(\zeta_0)$ and $w_2(\zeta_0)$, see Fig. 7.2 of~\cite{BMB}. For a reasonable value of $a_0$, such as a few parsecs, $\zeta_0 \sim 10^{-4}$ and $w$ is then very close to unity and we may neglect the contribution of $\mathcal{E}(r)$ to the effective dark matter. 

\subsection{NLG: Two-Body Force Law}

According to the Newtonian regime of NLG, the force of gravity on the point mass $m$ due to point mass $m'$ at position $\mathbf{r}$ is always attractive and is given by
\begin{equation}\label{I12}
 \mathbf{F}_{\rm NLG} =  \frac{Gmm' \mathbf{r}}{r^3} (1+ \Delta)\,, \qquad \Delta = -\mathcal{E}(r) +\alpha_0\,\left[1- (1+\tfrac{1}{2}\,\mu_0r)\,e^{-\mu_0r}\right]\,,
\end{equation}
where $r = |\mathbf{r}|$ and $\Delta$ is the net contribution of the effective dark matter. This force is conservative and satisfies Newton's third law of motion. With $a_0 = 0$,  $\mathcal{E}(r) = 0$ and the resulting force law has been employed in the gravitational physics of galaxies and clusters of galaxies to determine parameters $\alpha_0$ and $\mu_0$~\cite{Rahvar:2014yta}; indeed, observational data regarding the rotation curves of nearby spiral galaxies imply
\begin{equation}\label{I13}
\alpha_0 = 10.94 \pm 2.56\,,\quad \mu_0 = 0.059 \pm 0.028~{\rm kpc}^{-1}\,, \quad  \lambda_0 = \frac{2}{\alpha_0\,\mu_0} \approx 3\pm 2~{\rm kpc}\,.
\end{equation}
Furthermore, observational data regarding the solar system indicate that $a_0 > 10^{14}$~cm~\cite{Chicone:2015coa}. 

\begin{figure}
\centering
\includegraphics[width=8.0cm]{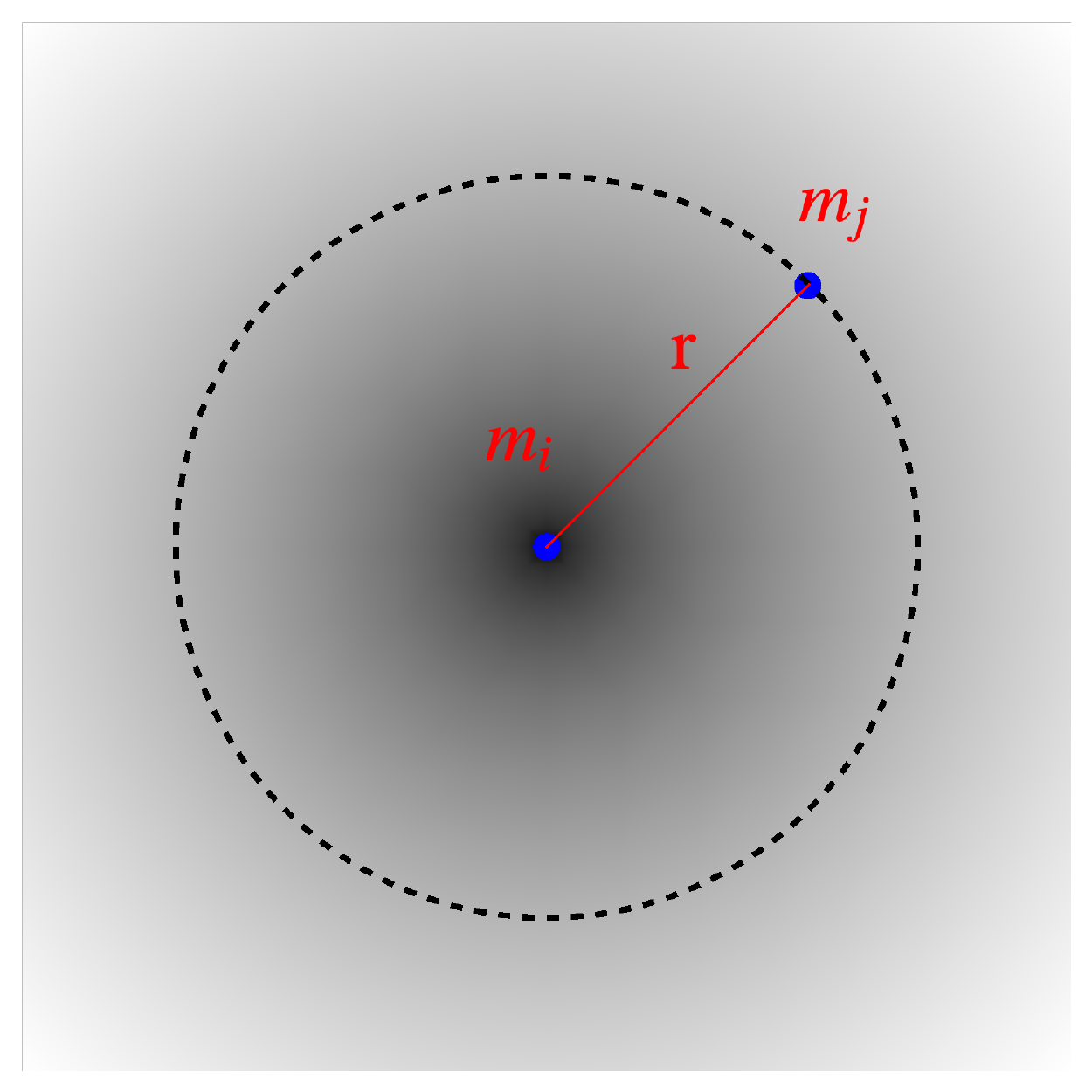}
\caption{Schematic diagram illustrates the calculation of the gravitational force on $m_j$  due to the effective dark mass associated with $m_i$ presented in Eq.~\eqref{I15}. Dark matter is indicated here by the spherically symmetric dark region that surrounds $m_i$ and extends to infinity. One can think of the effective dark mass inside the sphere of radius $r$ as though it were all concentrated at its center at $m_i$ due to Newton's shell theorem. Moreover, this theorem implies that the contribution of the effective dark mass outside the sphere of radius $r$ to the gravitational force on $m_j$ vanishes.}\label{fig1}
\end{figure}

It proves useful to present a physical derivation of the formula for the gravitational force~\eqref{I12}. In a system of $N$ point particles, the force on any particle $m_j$ is given by Eq.~\eqref{I12}, namely,
\begin{equation}\label{I14}
\mathbf{F}_{\rm NLG}^j = - Gm_j\sum_{i \ne j}^{N} \frac{m_i \mathbf{r}}{r^3} \left\{1- \mathcal{E}(r)+\alpha_0\,[1- (1+\tfrac{1}{2}\,\mu_0r)\,e^{-\mu_0r}] \right\}\,,
\end{equation}
where $\mathbf{r} = \mathbf{r}_i - \mathbf{r}_j$. This result consists of two terms: the Newtonian term due to $m_i$ plus the modification due to the fact that the point particle of mass $m_i$ is surrounded by a spherically symmetric distribution of effective dark matter of density $m_i\,q(r)$, where $q$ is the reciprocal kernel and its spherical symmetry is just a convenient  simplifying assumption. However, this simplifying assumption has the important result that by Newton's shell theorem the net attractive force of the dark matter associated with this spherical distribution points in the direction of $m_i$ and involves only that part of the spherical distribution of dark matter that is inside the radius $r = |\mathbf{r}_i - \mathbf{r}_j|$. To emphasize this point, the contributions of the Newtonian and effective dark matter parts can be written as 
\begin{equation}\label{I15}
 \frac{m_i}{r^2} + \frac{m_i \,\int_0^r 4\,\pi\,s^2 q(s) ds}{r^2}\,,
\end{equation}
where the rest of the spherical distribution of effective dark matter associated with $m_i$ does not contribute due to Newton's shell theorem. This is illustrated in Figure \ref{fig1}.

From Eq.~\eqref{I7}, we have
\begin{equation}\label{I16}
 \int_0^r 4\,\pi\,s^2 q(s) ds = -\mathcal{E} +  \int_0^r 4\,\pi\,s^2 q_0(s) ds\,,
\end{equation}
where, using Eq.~\eqref{I6}, we find
\begin{equation}\label{I17}
 \int_0^r 4\,\pi\,s^2 q_0(s) ds = \alpha_0\,\left[1- (1+\tfrac{1}{2}\,\mu_0r)\,e^{-\mu_0r}\right]\,.
\end{equation}
In this way,  we recover Eq.~\eqref{I14} for $\mathbf{F}_{\rm NLG}^j $ by summing over $m_i$, $i\ne j$.

\section{Effective Dark Matter}

It follows from Eq.~\eqref{I3} that in the Newtonian regime of NLG, Newtonian gravitation theory can be employed provided we take due account of the corresponding effective dark matter as well. Indeed,  every  point particle of mass $m$ is surrounded by a spherically symmetric distribution of effective dark matter of density $m\,q(r)$, where $r$ is the radial coordinate in a spherical polar coordinate system centered on the point particle $m$.  The net amount of effective dark matter is $m_D  = m\,\alpha_0\,w$, where $\alpha_0 \approx 11$. If we choose $q = q_0$, then $w = 1$. The result is slightly less if we choose $q_1$ or $q_2$, see Fig. 7.2 of~\cite{BMB}.
In any case, because of spherical symmetry, the center of mass of the \emph{total} effective dark matter associated with $m$ is exactly at the location of $m$. 

If we now have a Newtonian system of $N$ baryons, say,  each of inertial mass $m_i$, $ i = 1, 2, \cdots, N$, then each baryon brings with it the accompanying spherically symmetric effective dark matter of mass $m_i\,\alpha_0\,w$ centered on the baryon. Since $\alpha_0\,w$ is a constant and the same for each baryon, the Newtonian center of mass coincides with the center of mass of the \emph{total} effective dark matter. This will not be true, however, if we ask for the effective dark matter \emph{inside} a galaxy, for example;  then, the range of integration of the effective dark matter will be restricted by the boundary of the galaxy  and instead of $\alpha_0\,w$ for each baryon, we will get a fraction of this quantity depending on the location of the baryon $m_i$ in the galactic system.  Therefore, for a galaxy the baryonic center of mass is in general different from the center of mass of the corresponding effective dark matter. 

For any isolated system of $N$ baryons, we have the \emph{total} baryonic mass $M_B$ and the corresponding \emph{total} effective dark matter $\alpha_0\,w M_B$ over all space, where $w <1$. However, the amount of dark matter, $M_{D}$, confined within the system depends on the shape and volume of the system such that 
\begin{equation}\label{E1}
 M_{D}  < \alpha_0 \,w M_B <  \alpha_0 \, M_B\,, \qquad M_B := \sum_{i=1}^{N} \,m_i\,.
\end{equation}
The dynamic mass of the system is thus given by $M_B + M_{D} = M_B (1+\phi)$, where $\phi$ for the system is defined by  
\begin{equation}\label{E2}
\phi := \frac{M_{D}}{M_B} < \alpha_0\,w < \alpha_0\,.   
\end{equation}
In NLG, we have the basic result that \emph{for any finite system} the fraction $\phi$ must be smaller than $\alpha_0 \approx 11$. If the system is so huge that most of the baryonic matter is at distances larger than $\mu_{0}^{-1} \approx 17$ kpc, then we expect that the effective dark matter fraction approaches the limiting value of $\alpha_0\,w$.  Moreover, very far away from a baryonic system of mass $M_B$, it exhibits an  effective gravitational mass that approaches $M_B \, (1+ \alpha_0\,w)$. Therefore, for a cluster of galaxies, NLG predicts that we should have $\phi \approx \alpha_0\,w$. This appears to be consistent with observation of nearby clusters of galaxies~\cite{Rahvar:2014yta}. The standard definition of \emph{dark matter fraction} in astrophysics is
\begin{equation}\label{E3}
f_{DM} := \frac{M_{D}}{M_B + M_D} = \frac{\phi}{ 1 + \phi} < \phi\,.   
\end{equation}
Let us note that $f_{DM}$ as a function of $\phi$ increases monotonically from zero and approaches unity as $\phi \to \infty$. In NLG, $\phi < \alpha_0$; hence,  $f_{DM}$ is less than $\alpha_0/(1+\alpha_0)$. Therefore, for $\alpha_0 \approx11$, the effective dark matter ratio in NLG is such that  $f_{DM} < 0.917$. 

These considerations are naturally reflected in the gravitational force~\eqref{I12} on a point mass $m$ due to a point mass $m'$ located at $\mathbf{r}$.  For $r \gg 1/\mu_0$, Eq.~\eqref{I12} reduces to 
\begin{equation}\label{E4}
 \mathbf{F}_{\rm NLG} \approx  \frac{Gmm' \mathbf{r}}{r^3}(1+\alpha_0\,w)\,.
\end{equation}
That is, as the distance $r$ between $m$ and $m'$ increases, the magnitude of $\mathbf{F}_{\rm NLG}$ in Eq.~\eqref{I12} slowly increases from the Newtonian $1/r^2$ and for $r\gg\frac{1}{\mu_0}\approx 17$ kpc becomes $1/r^2$ again, but with $m' \to m'\,(1+\alpha_0\,w)$. The physical interpretation of this result is that $m$ is now subject to the attractive Newtonian gravitational force of $m$ as well as the total spherically symmetric effective dark matter distribution associated with $m'$. For $r \to \infty$, this net amount of dark matter becomes $m'\,\alpha_0\,w$, as expected, and hence the effective gravitational mass of $m'$ becomes $m'\,(1+\alpha_0 w)$. As is evident in the force formula above, this situation is symmetric between $m$ and $m'$ and hence our discussion could be repeated for the gravitational force experienced by $m'$. 

It is interesting to mention an alternative explanation of the behavior of the two-body gravitational force that is important for the consideration of dynamical friction that is the focus of the present work. We can imagine that the strength of the gravitational coupling increases with distance in NLG; that is, 
\begin{equation}\label{E4a}
G \to G ( 1 + \Delta)\,, \qquad \Delta =  1 + \int_0^r 4\,\pi\,s^2 q(s) ds\,, 
\end{equation}
so that the gravitational coupling ``constant" ranges from $G$ to $G(1+\alpha_0\,w)$ as $r$ goes from $0$ to $\infty$.

Let us briefly digress here and mention the connection of these results with Fourier space. From 
\begin{equation}\label{E5}
\hat{\rho} (\mathbf{k}) =  \int \rho(\mathbf{x})\, e^{-i\,\mathbf{k} \cdot \mathbf{x}}\, d^3x\,,
\end{equation}
we note that $\hat{\rho}(0) = M_B$ and similarly $\hat{\rho}_D(0)$ is the \emph{total} mass of dark matter over all space, namely, $\alpha_0\,w\,M_B$. The convolution theorem implies $\hat{\rho}_D(0) = \hat{q}(0)\, \hat{\rho}(0)$, which is consistent with the fact that $\hat{q}(0)= \alpha_0\,w$. Moreover, $\hat{q}_0(0) >  \hat{q}_1(0) >  \hat{q}_2(0)$, a relation that actually holds for any $\mathbf{k}$~\cite{BMB}.

Finally, let us imagine an isolated astrophysical system (``galaxy") consisting of $N$ particles and let $D$ be the baryonic diameter of the system, namely, the diameter of the smallest sphere that completely surrounds the baryonic system at the present time. For each baryon $m_i$ in the system, $i = 1, 2, \cdots, N$, the corresponding effective dark matter that is within the confines of the system and contributes to the dynamic mass of the whole system is evidently less than 
\begin{equation}\label{E6}
m_i\, \int_0^D 4\,\pi\,s^2 q(s) ds\,.
\end{equation}
This means that the total effective dark matter in the galaxy, $M_{D}$, is such that 
\begin{equation}\label{E7}
M_{D} < M_B\,  \int_0^D 4\,\pi\,s^2 q(s) ds\,. 
\end{equation}
On the other hand, $q < q_0$ implies 
\begin{equation}\label{E8}
 \int_0^D 4\,\pi\,s^2 q(s) ds \le \int_0^D 4\,\pi\,s^2 q_0(s) ds\,. 
\end{equation}
Moreover, Eq.~\eqref{I17} can be written as
\begin{equation}\label{E9}
\int_0^D 4\,\pi\,s^2 q_0(s) ds =  \alpha_0 \, H(D)\,,
\end{equation}
where
\begin{equation}\label{E10}
H(D) := 1- (1+\frac{1}{2} \mu_0\,D)\,e^{-\mu_0\,D}\,
\end{equation}
is a function that starts  from $H(0) = 0$ at $D=0$, monotonically increases with increasing $D$ and asymptotically approaches unity as $D \to \infty$. It follows that in the Newtonian regime of NLG, we have
\begin{equation}\label{E11}
\phi := \frac{M_{D}}{M_B} < \alpha_0 \, H(D) = \alpha_0 \,\left[1- (1+\tfrac{1}{2} \mu_0\,D)\,e^{-\mu_0\,D}\right]\,.
\end{equation}
For systems with $D \gg 1/\mu_0 \approx 17\,$kpc, such as for nearby clusters of galaxies, $H(D) \approx 1$ and we find
\begin{equation}\label{E12}
\phi \lessapprox \alpha_0\,,
\end{equation}
in agreement with the relation $\phi \approx \alpha_0\,w$ discussed earlier for such systems. On the other hand, for systems in the intermediate regime where $D$ is less than or comparable to  $1/\mu_0 \approx 17\,$kpc,  we note that $H(D)$ always stays below the line $\mu_0\,D/2$ for $D > 0$. Hence,  
\begin{equation}\label{E13}
\phi < \frac{D}{\lambda_0}\,,
\end{equation}
where $\lambda_0 \approx 3$ kpc. Thus for a globular star cluster, the effective amount of dark matter is at most a few percent of the baryonic content of the globular cluster. For a dwarf galaxy, the effective amount of dark matter could be at most comparable to its baryonic content. For giant galaxies and clusters of galaxies, $\phi$ can at most approach $\alpha_0\,w$. Further discussion of these issues is contained in~\cite{BMB}. For recent observational advances in connection with dwarf galaxies, see~\cite{vanDokkum:2018vup, Guo:2019wgb, Pina:2019rer, Hammer:2020qcd, Shen:2021zka} and the references cited therein; the prediction of NLG that there is less dark matter in dwarf galaxies seems to be consistent with these studies.

\section{Gravitational Drag}

The purpose of this section is to discuss the gravitational two-body system in accordance with the Newtonian limit of NLG in order to derive the analogue of Chandrasekhar's formula for dynamical friction. For the sake of simplicity and convenience, we follow closely the standard treatment presented in~\cite{Binney}.  An astronomical body of mass $M$ moves without direct collision through a population of stars, each with a typical mass $m$.  The result of two-body interaction of $M$ with a star of mass $m$ will then be extended to include all of the stars. 

\subsection{Two-Body Problem in NLG}

Imagine the ``Newtonian" gravitational interaction between masses $m$ and $M$ according to NLG. The main equations are
\begin{equation}\label{K1} 
m\,\frac{d^2\mathbf{x}_m}{dt^2}= \frac{GmM \mathbf{r}}{r^3}\,[1+ \Delta (r)]\,, \qquad M\,\frac{d^2\mathbf{x}_M}{dt^2}= - \frac{GmM \mathbf{r}}{r^3}\,[1+ \Delta (r)]\,,
\end{equation}
where $\Delta(r)$ is given by Eq.~\eqref{I12} and
\begin{equation}\label{K2} 
 \mathbf{r}:= \mathbf{x}_M - \mathbf{x}_m\,, \qquad r := |\mathbf{r}|\,.  
\end{equation}
As in the Newtonian Kepler problem, the center of mass moves uniformly and it is possible to separate the center of mass motion from the relative motion, so that 
\begin{equation}\label{K3} 
\ddot{\mathbf{r}}= - \frac{G(m+M) \mathbf{r}}{r^3}\,[1+ \Delta(r)]\,.  
\end{equation}
In NLG, $\Delta$ is a universal function of three new gravitational constants: $a_0$, $\alpha_0$ and $\mu_0$. Moreover, $\Delta(r)$ is a monotonically increasing function of $r$; indeed, $\Delta(r)$ starts from zero at $r = 0$ and eventually approaches $\alpha_0 w$ as $r\to \infty$. Equation~\eqref{K3} is rather complicated but possible to solve exactly; however, such a solution does not appear to be expressible in terms of the standard functions of mathematical physics and would hence be impractical for the problem under consideration in this paper. We therefore follow a different approach.   Let us note that in the classical two-body problem within the framework of NLG, the Newtonian gravitational constant $G$ is in effect replaced by $G(1+\Delta)$, indicating that the strength of the gravitational interaction increases by an order of magnitude in NLG as the relative distance goes to infinity. 
That is, for $r \ll \mu_0^{-1}$ and $r \gg \mu_0^{-1}$, we have the inverse-square law of gravity, while the gravitational coupling ``constant" ranges from $G$ to $G(1+ \alpha_0 w)$. It seems natural to employ an  approximation scheme involving a constant $\eta$ such that
\begin{equation}\label{K4} 
G(1+\Delta) \to  G(1+\eta) := G_{\eta}\,,  \qquad   0 < \eta < \alpha_0 w\,.
\end{equation}
The problem in NLG can thus be reduced in this approximation to the corresponding Newtonian problem; therefore, we can simply use the results of the standard treatment~\cite{Binney}, except that $G \to G_\eta=G(1+\eta)$. We defer the determination of $\eta$ to the end of this section.

\subsection{Dynamical Friction}

Next, we follow the standard treatment of this problem given in~\cite{Binney}. In the background inertial reference frame, mass $M$ with state $(\mathbf{x}_M, \mathbf{v}_M)$ interacts gravitationally with a star of mass $m$ with state $(\mathbf{x}_m, \mathbf{v}_m)$. Let $\mathbf{V}:=\dot{\mathbf{r}}$ be the relative velocity. The center of mass of the two-body system moves uniformly; therefore, the net change in the scattering process in the velocity of $M$ is given by $\delta \mathbf{v}_M = m \delta \mathbf{V}/(m+M)$ and, similarly, $\delta \mathbf{v}_m = -M \delta \mathbf{V}/(m+M)$.  In the standard scattering picture in terms of relative Cartesian coordinates, $\mathbf{V}_0$ is the initial relative velocity at $t = -\infty$. Hence, the magnitude of the relative specific orbital angular momentum is given by $L = b\,V_0$, where $b$ is the impact parameter. As a consequence of its distant encounter with a star of mass $m$, mass $M$ suffers a loss in its initial velocity along the original direction of motion due to the attractive drag of $m$.   We are interested in the net change in the velocity of $M$, $(\delta \mathbf{v}_M)_{||}$, along its initial direction of motion. The end result is
\begin{equation}\label{K5} 
(\delta \mathbf{v}_M)_{||} =  - 2\,m\,\mathbf{V}_0\,\frac{G_{\eta}^2 (m+M)}{G_{\eta}^2 (m+M)^2 + b^2 V_0^4}\,,  
\end{equation}
which is the same as given by~\cite{Binney}, except that we have replaced $G$ by $G_\eta = G(1+\eta)$, where $\eta$ is a constant parameter such $0 < \eta < \alpha_0 w$. The magnitude of $\eta$ for a given system is estimated at the end of this section. The presence of $\eta$ means that the net effect is naturally stronger in NLG due to the effective dark matter of $m$. 

The rest of the analysis is exactly as in Ref.~\cite{Binney} and we find the Chandrasekhar dynamical friction formula
\begin{equation}\label{K6} 
\frac{d \mathbf{v}_M}{dt} =  - 16 \pi^2 G_{\eta}^2 \,m\,(m+M)\ln \Lambda\,\frac{\mathbf{v}_M}{v_M^3}\,\int_0^{v_{M}} f(v) v^2 \,dv\,,  
\end{equation}
where $G$ has again been replaced by $G_\eta$ and $f(v)$ is the isotropic velocity distribution function of the background stars. Here, $\ln \Lambda$ is the Coulomb logarithm given by
\begin{equation}\label{K7} 
\Lambda =  \frac{b_{\rm max} V_0^2}{G_\eta (m + M)}\approx \frac{D\, v_{\text{typ}}^2}{G_\eta (m + M)}\,,  
\end{equation}
where $v_{\text{typ}}$ is the typical velocity of the background particles.  For $M \gg m$, the magnitude of dynamical friction force $\mathcal{F}$ can be expressed as
\begin{equation}\label{K7a}
\mathcal{F}_{\text{NLG}}=-\frac{4\pi\, G_{\eta}^2\, M^2}{v_{M}^2}\rho(<v_M)\ln\Lambda\,,
\end{equation}
where  $\rho$ is the mass density of the background stars and  
\begin{equation}\label{K7b}
\rho(<v_M)=4\pi \,m\int_0^{v_{M}} f(v) v^2 \,dv\,.
\end{equation}
It is possible to write $\Lambda = b_{\rm max}/b_{\rm min}$, where in the present approach $b_{\rm min} = G_\eta (m + M)/v_{\text{typ}}^2$. As revealed by simulations, Chandrasekhar's formula is quite useful in astrophysical applications~\cite{Binney}; nevertheless, it is an approximate result that takes into account only the two-body gravitational scatterings of $M$ with the stars of the background infinite homogeneous medium.    

It is important to compare our result, namely, $\mathcal{F}_{\text{NLG}}$ with the standard $\Lambda$CDM picture involving particles of dark matter. Let us first note that one can recover the Newtonian dynamical friction force, $\mathcal{F}_{\text{N}}$, by simply setting $\eta$ equal to zero. On the other hand, it would then be necessary to take into account the contribution of the hypothetical dark matter particles. We recall that such particles are postulated to be nonexistent in NLG. Assuming that dark matter particles exist with mass density $\rho_d$, $\mathcal{F}_{\text{N}}$ can be written as
\begin{equation}\label{K7c}
\mathcal{F}_{\text{N}}=-\frac{4\pi\, G^2\, M^2\, }{v_{M}^2}\Big[\rho(<v_M)\ln\tilde{\Lambda}+\rho_d(<v_M)\ln\tilde{\Lambda}_d\Big]\,.
\end{equation}
Here, $\ln \tilde{\Lambda}$ and $\ln \tilde{\Lambda}_d$ with   
\begin{equation}\label{K7d}
 \tilde{\Lambda}=\frac{D\, v_{\text{typ}}^2}{G (m + M)},\qquad  \tilde{\Lambda}_d=\frac{D_d\, v_{\text{typ}}^2}{G (m_d + M)}\,
\end{equation}
are the Coulomb logarithms associated with the baryonic and dark matter particles, respectively,  and $D_d$ is the diameter of the smallest sphere that completely surrounds the corresponding dark matter particles of the system at the present epoch. Comparing Eq.~\eqref{K7a} with Eq.~\eqref{K7c}, it turns out that depending on the  physical properties of the system, dynamical friction can be 
stronger, equivalent or weaker in NLG as compared to the Newtonian theory that includes dark matter particles. It appears that $|\mathcal{F}_{\text{NLG}}/\mathcal{F}_{\text{N}}|$ depends sensitively upon the ratio $\rho_d/\rho$. For instance, when $\rho_d < \rho$, it is possible that $|\mathcal{F}_{\text{NLG}}|$ is greater than, or equal to, $|\mathcal{F}_{\text{N}}|$. In the latter case, dynamical friction cannot be employed to distinguish between nonlocal gravity and Newtonian theory with particle dark matter. In Section V, we consider a certain astrophysical system explicitly and compare the role of dynamical friction in NLG with the standard Lambda cold dark matter (Lambda-CDM) case.

\subsection{Estimation of $\eta$}

To estimate $\eta$, the simplest possibility would be to average the function $\Delta(r)$ that appears in Eqs.~\eqref{I12} and~\eqref{K1} and represents the contribution of effective dark matter to the ``Newtonian" gravitational force. Let us recall here that $\Delta$ as defined by Eq.~\eqref{I12} consists of two parts. The term involving $\mathcal{E}$ vanishes in the absence of the short-distance parameter $a_0$. Indeed, for a reasonable value of $a_0$ --- for instance, for $a_0$ equal to a few parsecs --- the contribution of $\mathcal{E}(r)$ to  $\Delta(r)$ can be neglected. Then, 
\begin{equation}\label{K8} 
\Delta(r) \approx  \alpha_0\,H(r)\,,
\end{equation}
where $H$ is defined by Eq.~\eqref{E10}. From
\begin{equation}\label{K9} 
\int H(r) \,dr =  \alpha_0\,r  +\frac{1}{2}\, \frac{\alpha_0}{\mu_0}\, (3 + \mu_0\,r)\,e^{-\mu_0\,r}\,,
\end{equation}
we find that averaging $\Delta(r)$ from $r_1 $ to $r_2 $ results in
\begin{equation}\label{K10} 
\eta \approx \alpha_0\,(r_2-r_1)^{-1}\int_{r_1}^{r_2} H(r) \,dr =  \alpha_0  + \frac{\alpha_0}{2\mu_0(r_2-r_1)}\Big[ (3+\mu_0r_2)e^{-\mu_0\,r_2} - (3+\mu_0r_1)e^{-\mu_0\,r_1} \Big]\,.
\end{equation}
Here, $r_1 \approx b_{\rm min}$ could be considered to be the maximum of $G_\eta (m + M)/v_{\text{typ}}^2$ and the size of the reference body $M$, while $r_2 \approx b_{\rm max}$ could be taken to be the diameter of the baryonic system $D$. That is, as before,  $D$ is the diameter of the smallest sphere that completely surrounds the baryonic system (galaxy) at the present time. Let us recall that $\Lambda \approx r_2/r_1$ and assume for the sake of definiteness that
\begin{equation}\label{K11} 
r_2 = D\,, \qquad r_1 =\frac{D}{\Lambda}\,.
\end{equation}
Then, $\eta$ can be expressed as
\begin{equation}\label{K12}
\eta \approx  \alpha_0 + \frac{\alpha_0 \Lambda}{2(\Lambda -1) \mu_0 D} \left[(3+\mu_0 D)\,e^{-\mu_0 D} - (3+ \tfrac{\mu_0 D}{\Lambda})\,e^{-\frac{\mu_0 D}{\Lambda}}\right]\,.
\end{equation}
Normally, we have $r_2 \gg r_1$ (or equivalently $\Lambda\gg 1$) and $\mu_0\,r_1 \ll 1$. In fact, the limiting case of $\Lambda \to \infty$ exists and we find 
\begin{equation}\label{K13}
\eta (\Lambda \to \infty) \to \alpha_0\,\left(1 -\frac{3}{2}\,\frac{1-e^{-\mu_0\,D}}{\mu_0\,D} +\frac{1}{2}\,e^{-\mu_0\,D}\right)\,.
\end{equation}
For $\mu_0\,D \in (0, \infty)$, the function in the parentheses starts from zero with slope $1/4$ and slowly but monotonically approaches unity as $\mu_0\,D \to \infty$.  

To make a crude estimate for $\eta$ in a globular star cluster, a dwarf galaxy and a galaxy, let us take the following typical sizes $D\approx 10\,$pc, $\approx 1\,$kpc and $\approx 10\,$kpc, respectively. Using equation~\eqref{K12} with $\Lambda \approx 100$, $\eta$ for a globular star cluster is small, i.e. $\eta\approx 10^{-3}$, implying that effects of NLG do not appear to be significant in globular star clusters. On the other hand, for a dwarf galaxy and a normal galaxy we find $\eta\approx 0.163$ and $\eta\approx 1.596$, respectively.

\section{Gravitational Wake}

There is an alternative way to view Chandrasekhar's dynamical friction formula. As the reference body of mass $M$ moves through an infinite homogeneous medium consisting of stars of average mass $m$, the ensuing disturbance leads to a density enhancement in the reference body's wake, thereby slowing it down via gravitational attraction. This approach has been developed by a number of authors; see~\cite{TW} and the references cited therein. To determine the role of NLG in this process, we follow the treatment presented in~\cite{TW}. 

In the background medium, we ignore the gravitational interaction between the stars. Therefore, the stars in their unperturbed states move according to $\mathbf{x} = \mathbf{x}_0 + \mathbf{v}_m t$ with constant momenta $\mathbf{p} = m \mathbf{v}_m$ starting from $ t = -\infty$ with $\mathbf{x}_0$ ranging over all space. The motion of the reference body generates a density perturbation
\begin{equation}\label{W1}
\rho_M(t, \mathbf{x}) = M e^{\beta t} \delta(\mathbf{x} - \mathbf{v}_M t)\,,
\end{equation} 
where $\beta > 0$ is a constant auxiliary parameter that we need in the course of our calculations; however, we will eventually assume that $\beta \to 0$. In the Newtonian regime of NLG, Poisson's Eq.~\eqref{I3} takes the form 
\begin{equation}\label{W2}
\nabla^2\Phi = 4\pi GMe^{\beta t}\,[\delta(\mathbf{x} - \mathbf{v}_M t) + q(\mathbf{x} - \mathbf{v}_M t)]\,,
\end{equation} 
where $q$ is the reciprocal kernel. Working in Fourier space, we find
\begin{equation}\label{W3}
\Phi (t, \mathbf{x})  = -\frac{GM}{2\pi^2} \int \frac{1+\hat{q}(k)}{k^2}\, e^{i \mathbf{k}\cdot (\mathbf{x} - \mathbf{v}_M t) +\beta t} \,d^3k\,,
\end{equation} 
where we have used the fact that the reciprocal kernel is spherically symmetric by assumption and its Fourier integral transform $\hat{q}(k)$ is thus only a function of $k = |\mathbf{k}|$. Neglecting any interaction between the stars, the motion of a star is perturbed by the gravitational influence of the reference body alone. In general, 
\begin{equation}\label{W4}
\frac{d\mathbf{p}}{dt} = - m \nabla\Phi(\mathbf{x})\,.
\end{equation}
We can express the resulting perturbation in $\mathbf{x}$ and $\mathbf{p}$ of a star in powers series in terms of the gravitational coupling constant $G$. That is, 
\begin{equation}\label{W5}
\mathbf{x} = \mathbf{x}_0 + \mathbf{v}_m t + \Delta_1\mathbf{x} +   \Delta_2\mathbf{x} +\cdots\,, \qquad \mathbf{p} = m \mathbf{v}_m + \Delta_1\mathbf{p} +   \Delta_2\mathbf{p} +\cdots\,,
\end{equation}
where, for instance,  $ \Delta_n\mathbf{p}$ is the momentum perturbation of order $n$. Substituting Eq.~\eqref{W5} in Eq.~\eqref{W4}, the first-order perturbation in $\mathbf{p}$ can be obtained from Eq.~\eqref{W4} with unperturbed $\mathbf{x} = \mathbf{x}_0 + \mathbf{v}_m t$. Therefore,  
\begin{equation}\label{W6}
\frac{d (\Delta_1p^j)}{dt}  = i\frac{GMm}{2\pi^2} \int \frac{1+\hat{q}(k)}{k^2}\, k^j\,e^{i (-\omega t +\mathbf{k}\cdot \mathbf{x}_0)}\,d^3k\,,
\end{equation} 
where
\begin{equation}\label{W7}
\omega  := \mathbf{k}\cdot(\mathbf{v}_M - \mathbf{v}_m) + i \beta\,.
\end{equation} 
It is important to note that the net force of gravity on the stars vanishes to first order in $G$. For instance, the $j$-component of the net force can be calculated by integrating Eq.~\eqref{W6} over all the stars. To this end, let us multiply the right-hand side of Eq.~\eqref{W6} by $N\,d^3x_0$, where $N$ is the constant density of stars. The integration over the $j$-component of  $\mathbf{x}_0$ results in $\xi \,\delta(\xi) = 0$ in the integrand, where $\xi = k^j$. Thus, the net rate of momentum transfer is zero to linear order in $G$. We therefore proceed to the evaluation of this quantity to second order in $G$.  

The second-order perturbation in the momentum $\Delta_2\mathbf{p}$ can be determined from
\begin{equation}\label{W8}
\frac{d(\Delta_2 p_j)}{dt}  = - m \frac{\partial^2 \Phi}{\partial x^j \partial x^l} \,\Delta_1 x^l\,,
\end{equation} 
where $K_{jl} =  \partial^2 \Phi/\partial x^j \partial x^l$ is the tidal matrix evaluated along the unperturbed path of a star. Moreover, the first-order change in the position of a star can be calculated by integrating the force Eq.~\eqref{W6} twice over time starting from $t = -\infty$. The result is
\begin{equation}\label{W9}
\Delta_1 x^j  = -i\frac{GM}{2\pi^2} \int \frac{1+\hat{q}(k)}{k^2 \omega^2}\, k^j\,e^{i (-\omega t +\mathbf{k}\cdot \mathbf{x}_0)}\,d^3k\,.
\end{equation} 

To evaluate this second-order rate of momentum transfer, let us first note that 
\begin{equation}\label{W10}
\frac{\partial^2 \Phi}{\partial x^j \partial x^l} = \frac{GM}{2\pi^2} \int \frac{1+\hat{q}(k')}{k'^2}\, k'_jk'_l\,e^{i (-\omega' t +\mathbf{k'}\cdot \mathbf{x}_0)}\,d^3k'\,,
\end{equation} 
where $\omega'  := \mathbf{k'}\cdot(\mathbf{v}_M - \mathbf{v}_m) + i \beta$. It proves convenient at this point to replace $\mathbf{k'}$ by $-\mathbf{k'}$ and hence $\omega'$ by $-\omega^{'*}$ in Eq.~\eqref{W10}; then, combining Eqs.~\eqref{W9} and~\eqref{W10} we find
\begin{equation}\label{W11}
\frac{d(\Delta_2 p^j)}{dt}  = i\, m\left (\frac{GM}{2\pi^2}\right)^2 \int \frac{[1+\hat{q}(k)][1+\hat{q}(k')]}{k^2k'^2\omega^2}\, k'^jk'_lk^l\,e^{i (\omega^{'*} -\omega) t +i(\mathbf{k}-\mathbf{k'})\cdot \mathbf{x}_0}\,d^3k\,d^3k'\,.
\end{equation} 
To find the net rate of momentum transfer to second order, we must sum over all the stars; therefore, let us integrate Eq.~\eqref{W11} over $N\,d^3x_0$. The result is
\begin{equation}\label{W12}
\Sigma_{\rm stars} \frac{d(\Delta_2 p^j)}{dt}  =\frac{2 i}{\pi} G^2 M^2 m Ne^{2\beta t}\int \frac{[1+\hat{q}(k)]^2}{k^2\omega^2}\, k^j\,d^3k\,,
\end{equation} 
where we have used the fact that $i (\omega^{'*} -\omega) = 2 \beta$ when $\mathbf{k'} = \mathbf{k}$. In Appendix A, we prove a useful identity, namely, 
\begin{equation}\label{W13}
i \int \frac{[1+\hat{q}(k)]^2}{k^2\omega^2}\, k_j\,d^3k = \beta \frac{\partial}{\partial v_m^j} \int \frac{[1+\hat{q}(k)]^2}{k^2|\omega|^2}\,d^3k\,.
\end{equation} 
Substituting this relation in Eq.~\eqref{W12} and taking advantage of the following representation of Dirac's delta function: 
\begin{equation}\label{W14}
\delta (x)  = \lim_{\beta \to 0^{+}} \frac{1}{\pi} \frac{\beta}{|x+ i\beta|^2}\,,
\end{equation} 
we find by letting $\beta \to 0^{+}$ that $\exp(2 \beta t) \to 1$ and
\begin{equation}\label{W15}
\Sigma_{\rm stars} \frac{d(\Delta_2 p_j)}{dt}  = 2 G^2 M^2 m N  \frac{\partial}{\partial v_m^j}\int \frac{[1+\hat{q}(k)]^2}{k^2}\, \delta[\mathbf{k}\cdot(\mathbf{v}_M - \mathbf{v}_m)]\,d^3k\,.
\end{equation} 

Thus far, we have been working with Cartesian coordinates in Fourier space. Let us now introduce spherical polar coordinates and write
\begin{equation}\label{W16}
k_1 = k \sin\vartheta \cos\varphi\,, \qquad  k_2 = k \sin\vartheta \sin\varphi\,,\qquad k_3 = k\cos\vartheta\,.
\end{equation} 
Without any loss in generality, we can choose the Cartesian axes in Fourier space such that $\mathbf{v}_M - \mathbf{v}_m$ is a vector in the polar direction; then, integrating over angles in Eq.~\eqref{W15}, we get
\begin{equation}\label{W17}
\Sigma_{\rm stars} \frac{d(\Delta_2 p_j)}{dt}  = 4\pi G^2 M^2 m N  \frac{\partial}{\partial v_m^j}\frac{1}{|\mathbf{v}_M - \mathbf{v}_m|}\int [1+\hat{q}(k)]^2\frac{dk}{k}\,.
\end{equation}
It follows from the results of Appendix A and Eq.~\eqref{W20} below that this net force is in the direction of  $\mathbf{v}_M$, once we take due account of the isotropic velocity distribution of the background stars. 

We note that the mass density of stars is given by
\begin{equation}\label{W18}
\rho = m N = m \int f(v_m) d^3v_m\,,
\end{equation} 
where $f(v_m)$ is the isotropic velocity distribution function of the stars. Therefore, in Eq.~\eqref{W17}, the part that depends on the velocity of the stars should be replaced by 
\begin{equation}\label{W19}
 m N \frac{\partial}{\partial v_m^j}\frac{1}{|\mathbf{v}_M - \mathbf{v}_m|} \mapsto m \int f(v_m) \frac{(\mathbf{v}_M - \mathbf{v}_m)_j}{|\mathbf{v}_M - \mathbf{v}_m|^3}\,d^3v_m\,,
\end{equation} 
since Eq.~\eqref{W17} involves the sum over all stars and that includes their velocity distribution. Using Newton's shell theorem in the present context, we find
\begin{equation}\label{W20}
  \int f(v_m) \frac{(\mathbf{v}_M - \mathbf{v}_m)}{|\mathbf{v}_M - \mathbf{v}_m|^3}\,d^3v_m  = 4\pi \frac{\mathbf{v}_M}{v_M^3} \int_0^{v_M} f(v_m) v_m^2\, dv_m\,. 
\end{equation} 
Finally, Eq.~\eqref{W17} can be written as
\begin{equation}\label{W21}
\Sigma_{\rm stars} \frac{d(\Delta_2 p^j)}{dt}  = 16\pi^2 G^2 M^2 m \frac{v_M^j}{v_M^3} \int_0^{v_M} f(v_m) v_m^2\, dv_m\int [1+\hat{q}(k)]^2\frac{dk}{k}\,,
\end{equation} 
where 
\begin{equation}\label{W22}
\rho(< v_M)   = 4\pi m \int_0^{v_M} f(v_m) v_m^2\, dv_m\,.
\end{equation} 

The total rate of momentum transfer to the stars~\eqref{W21} should be equal and opposite to the dynamical friction force $\mathcal{F}$ experienced by the reference body; hence, 
\begin{equation}\label{W23}
\mathcal{F}_{\rm NLG} = M \frac{d\mathbf{v}_M}{dt}  = -4\pi G^2 M^2 \frac{\mathbf{v}_M}{v_M^3} \rho(< v_M)\int [1+\hat{q}(k)]^2\frac{dk}{k}\,,
\end{equation} 
which reduces to Chandrasekhar's result in the absence of the reciprocal kernel. That is, the Coulomb logarithm is given by 
\begin{equation}\label{W24}
\int_{k_{\rm min}}^{k_{\rm max}} \frac{dk}{k} = \ln \left(\frac{k_{\rm max}}{k_{\rm min}}\right)\,,
\end{equation} 
where $1/k_{\rm max}$ has to do with the distance of closest approach $(``b_{\rm min}")$, while $1/k_{\rm min}$ has to do with the extent of the background medium $(``b_{\rm max}")$ which we assumed to be equal to $D$. 

To connect with our approach in the previous section, we can define a constant parameter $\hat{\eta}$ via
\begin{equation}\label{W25}
\int_{k_{\rm min}}^{k_{\rm max}} [1+\hat{q}(k)]^2\frac{dk}{k} = (1+\hat{\eta})^2 \int_{k_{\rm min}}^{k_{\rm max}} \frac{dk}{k}\,. 
\end{equation} 
Then, Eq.~\eqref{W23} takes the form of Chandrasekhar's formula with $G$ replaced by $G_{\hat \eta} = G (1+\hat{\eta})$, where $\hat{\eta}$ depends upon the boundaries of the domain of integration $k_{\rm min}$ and $k_{\rm max}$; that is, 
\begin{equation}\label{W26}
\mathcal{F}_{\rm NLG} = M \frac{d\mathbf{v}_M}{dt}  = -4\pi G_{\hat{\eta}}^2 M^2 \frac{\mathbf{v}_M}{v_M^3} \rho(< v_M) \ln \Lambda\,,
\end{equation}
where $\Lambda =  b_{\rm max} / b_{\rm min}$.  

\subsection{Estimation of $\hat{\eta}$}

The reciprocal kernel depends upon three parameters, namely, $a_0$, $\lambda_0$ and $\mu_0$. To estimate $\hat{\eta}$, we neglect the short-distance parameter $a_0$ for the sake of simplicity.  In this case, $q$ reduces to $q_0$ given by Eq.~\eqref{I6}. Therefore, 
\begin{equation}\label{E0}
\int_{k_{\rm min}}^{k_{\rm max}} [1+\hat{q}_0(k)]^2\frac{dk}{k} \approx (1+\hat{\eta})^2 \int_{k_{\rm min}}^{k_{\rm max}} \frac{dk}{k}\,, 
\end{equation} 
where  $\hat{q}_0(k)$ is the Fourier integral transform of $q_0$ and is given by
\begin{equation}\label{E1}
\hat{q}_0(k)=\frac{\alpha_0}{2}\Big(\frac{1}{1+k^2/\mu_0^2}+\frac{\mu_0}{k}\arctan\frac{k}{\mu_0}\Big)>0\,.
\end{equation}
Notice that $\hat{q}_0$ goes from $\alpha_0$ to 0 when $k$ goes from 0 to $\infty$; therefore, we expect from Eq.~\eqref{E0} that  $0 < \hat{\eta} < \alpha_0$. It is straightforward to evaluate the integral containing $\hat{q}_0$ in Eq.~\eqref{E0} in order to find an approximate analytic expression for $\hat{\eta}$. Let us define
\begin{equation}\label{E2}
\mathcal{I} := \int[1+\hat{q}_0(k)]^2\frac{dk}{k} =\int\Big[1+\frac{\alpha_0}{2}\Big(\frac{1}{1+u^2}+\frac{\arctan u}{u}\Big)\Big]^2\frac{d u}{u}\,,
\end{equation}
where $u=k/\mu_0$. This integral can be evaluated explicitly and the result is 
\begin{align}\label{E3}
\mathcal{I} (u) = {}&\frac{\alpha_0 ^2}{8} \frac{1}{u^2+1} + (\alpha_0 +1)^2 \ln u -\frac{1}{2} \alpha_0 (\alpha_0 +2) \ln(u^2 + 1)    \\ 
\nonumber {}&- \frac{\alpha_0}{4}(3\alpha_0 + 4)\frac{\arctan u}{u} - \frac{\alpha_0 ^2}{8}(3u^2+1) \frac{\arctan^2 u}{u^2}\,.
\end{align}
We recall that $k_{\rm min} = 1/ b_{\rm max} \approx 1/D$ and $k_{\rm max} = 1/  b_{\rm min} \approx \Lambda/D$. With $u=k/\mu_0$,
\begin{equation}\label{E4}
 u_{\rm min} \approx \frac{1}{\mu_0 D}\,, \qquad u_{\rm max} \approx \frac{\Lambda}{\mu_0 D}\,,
\end{equation}
as before, we can evaluate $\hat{\eta}$ from
\begin{equation}\label{E5}
\mathcal{I} (u_{\rm max}) - \mathcal{I} (u_{\rm min}) \approx (1+ \hat{\eta})^2 \ln\left(\frac{u_{\rm max}}{u_{\rm min}}\right)\,. 
\end{equation}
It is interesting to note that $\hat{\eta} (\Lambda \to \infty) \to 0$, which should be compared and contrasted with the corresponding Eq.~\eqref{K13} for $\eta$. 

To estimate $\hat{\eta}$, which is in some sense the Fourier analogue of $\eta$ discussed at the end of Section III, we use $\Lambda \approx 100$, as before, and the same typical values for $D$ used to estimate $\eta$, namely, $\approx 10$ pc, $\approx 1\,$kpc and $\approx 10\,$kpc for a globular cluster, a dwarf galaxy and a galaxy, and the corresponding estimates for $\hat{\eta}$ turn out to be $\approx10^{-3}$, $\approx 0.116$ and $\approx 1.409$, respectively.  It is clear that our estimates for $\hat{\eta}$ are in fairly good agreement with those for $\eta$.

\section{Dynamical Friction in Barred Spiral Galaxies}

In the standard $\Lambda$CDM model,  dynamical friction plays a central role in the secular evolution of spiral galaxies. In particular, a stellar bar in a spiral galaxy transfers angular momentum between the inner and outer parts of the galactic disk. More importantly, the stellar bar exchanges angular momentum with the dark matter halo due to dynamical friction on the stellar bar in its interaction with the particles of dark matter. In comparison, the corresponding dynamical friction on the stellar bar due to its interaction with the baryonic content in the galactic disk is negligibly small. One may simply regard the first term in Eq.~\eqref{K7c} to be negligible in comparison with the second term. That is, the wake induced by the motion of the bar in the normal disk particles is much weaker than the corresponding wake induced in the particles of dark matter. 

The angular momentum exchange via dynamical friction means that the pattern speed of the bar decreases with time~\cite{Debattista:2000ey}. This circumstance implies that barred spiral galaxies in $\Lambda$CDM model should host relatively slow bars, as indeed occurs in almost all of the state-of-the-art cosmological hydrodynamic simulations~\citep{Roshan:2021liy}. On the other hand, most of the observed bars are relatively fast, which has posed a challenge for the standard $\Lambda$CDM model. One possibility involves postulating \emph{ultralight} axions as the particles of dark matter~\cite{Hui:2016ltb}. One of the advantages of such \emph{fuzzy} dark matter particles is that dynamical friction can be substantially suppressed in spiral galaxies. However, a recent study of such models claims that these axions are in gross disagreement with Lyman-alpha forest observations at 99.7\% confidence level~\cite{Rogers:2020ltq, Rogers:2020cup}.  

In the context of NLG, there is no dark matter halo. Depending on the value of $\hat{\eta}$, we can quantify the impact of dynamical friction on the stellar bar due to the baryonic content of the disk. As already mentioned,  this makes a negligible contribution to the total dynamical friction in the standard $\Lambda$CDM picture. It is necessary to mention that our result for dynamical friction, i.e. Eqs.~\eqref{K7a} and~\eqref{W26}, have been obtained for a moving point mass and not a stellar bar. However, modeling a bar with a dumbbell consisting of two point masses captures some important features of a more detailed description~\cite{MDW, Binney}. To estimate $\hat{\eta}$ for the apparent rotation of a stellar bar parallel to the disk of a spiral galaxy, it is appropriate to take $b_{\rm max} \approx h$, where $h$ is the thickness of the disk. The typical value for $h$ is $\approx 2\,$kpc. In this case, we find  $\hat{\eta}\approx 0.244$ using Eq.~\eqref{E5}. Therefore, the contribution of the baryonic matter to the dynamical friction should still be essentially negligible even though it is stronger in NLG by a factor of $(1+\hat{\eta})^2 \approx 1.55$. That is, it is rather unlikely that this 55\% increase in dynamical friction by baryonic matter in NLG can mimic in magnitude the strong contribution of the dark matter particles in the standard $\Lambda$CDM picture. The numerical galactic simulations in NLG confirm this description; that is,  for all of the maximal disks simulated in~\cite{Roshan:2019xda, Roshan:2021mfc}, dynamical friction is essentially negligible in the NLG disks, while a substantial reduction in the bar speed appears in the standard $\Lambda$CDM dark matter models.

\section{Discussion}

Chandrasekhar's dynamical friction appears as an important aspect of astrophysical systems helping to distinguish between particle dark matter and modified gravity. We have therefore studied dynamical friction in the Newtonian regime of nonlocal gravity (NLG). Within the context of NLG, we have followed the standard approaches to dynamical friction and demonstrated that nonlocal gravity effects appear effectively as an enhancement in the strength of Newton's gravitational constant $G$. Specifically, we have shown that Chandrasekhar's expression holds approximately in NLG as well and the only essential difference is that $G$ is replaced by $G\,(1+\hat{\eta})$, where $\hat{\eta}$ is a constant such that $0<\hat{\eta}<\alpha_0 \approx 11$. This means that in systems where there is no actual dark matter component, NLG leads to stronger dynamical friction. On the other hand, in systems hosting a substantial amount of hypothetical dark matter, depending on the magnitude of $\hat{\eta}$ and the other physical properties, dynamical friction can be stronger or weaker than in NLG. As an instance of possible astrophysical implications of our results, we have argued that, in general agreement with numerical simulations~\cite{Roshan:2019xda, Roshan:2021mfc}, dynamical friction on stellar bars in barred galaxies is much weaker in NLG compared to the standard $\Lambda$CDM picture.

\section*{Acknowledgments}
 
The work of M.R. has been supported by the Ferdowsi University of Mashhad.

\appendix

\section{Proof of  Identity~\eqref{W13}}

Employing spherical polar coordinates in Fourier space introduced in Eq.~\eqref{W16}, we can write the left-hand side of identity~\eqref{W13} as
\begin{equation}\label{A1}
\mathbb{I}_j = i \int \frac{[1+\hat{q}(k)]^2}{k^2\omega^2}\, k_j\,d^3k =  i \int \frac{[1+\hat{q}(k)]^2}{(kU \cos\vartheta + i \beta)^2}\,k_j  \sin\vartheta\,dk d\vartheta d\varphi\,,
\end{equation}
where $U := |\mathbf{v}_M - \mathbf{v}_m|$ and $\mathbf{v}_M - \mathbf{v}_m$ has been assumed to be a vector in the polar direction with no loss in generality. Integrating over the azimuthal angle $\varphi$, we find $\mathbb{I}_1= \mathbb{I}_2 = 0$ for $j=1$ and $j=2$. For $j = 3$, we have  
\begin{equation}\label{A2}
\mathbb{I}_3 = 2\pi i \int \frac{[1+\hat{q}(k)]^2}{(k^2U^2 \zeta^2 +\beta^2)^2} (kU\zeta - i \beta)^2 k\zeta\, dk d\zeta\,,
\end{equation}
where $\zeta = \cos\vartheta \in (-1, 1)$. Thus, only even terms in $\zeta$ contribute to the integrand for $\mathbb{I}_3$ and we have 
\begin{equation}\label{A3}
\mathbb{I}_3 = 4\pi \beta U\int \frac{[1+\hat{q}(k)]^2}{(k^2U^2 \zeta^2 +\beta^2)^2} k^2\zeta^2\, dk d\zeta\,.
\end{equation}

Let us consider next the right-hand side of identity~\eqref{W13},
\begin{equation}\label{A4}
\beta \frac{\partial}{\partial v_m^j} \int \frac{[1+\hat{q}(k)]^2}{k^2|\omega|^2}\,d^3k\,
\end{equation} 
and note that 
\begin{equation}\label{A5}
\frac{\partial}{\partial v_m^j}\,\frac{1}{|\omega|^2} = \frac{\partial}{\partial v_m^j}\,\frac{1}{[\mathbf{k}\cdot(\mathbf{v}_M - \mathbf{v}_m)]^2 +  \beta^2} = 2 kU \zeta \frac{k_j}{(k^2U^2 \zeta^2 +\beta^2)^2}\,.
\end{equation} 
Substituting Eq.~\eqref{A5} in Eq.~\eqref{A4}, we find
\begin{equation}\label{A6}
2\beta U \int \frac{[1+\hat{q}(k)]^2}{(k^2U^2 \zeta^2 +\beta^2)^2} k_j k \zeta\, dk d\zeta d\varphi\,,
\end{equation} 
which vanishes for $j = 1$ and $j = 2$, while for $j = 3$ integrates over the azimuthal angle $\varphi$ to the same result as $\mathbb{I}_3$. This completes the proof of  identity~\eqref{W13}.

\end{document}